\documentclass[aps,prd,preprint,superscriptaddress,nofootinbib,showkeys]{revtex4}
\usepackage{graphics}
\usepackage{epsfig}
\usepackage{latexsym}
\usepackage{colordvi}
\usepackage{amsmath}
\usepackage{amssymb}

\newcommand{\be}{\begin{equation}}
\newcommand{\ee}{\end{equation}}

\begin{document}
\preprint{\parbox[b]{1in}{ \hbox{\tt PNUTP-09-A01}  }}

\title{Baryons in dense QCD~\footnote{A contribution to  ``Multifaceted Skyrmion", edited by
G. Brown and M. Rho, World Scientific.}}

\author{Deog Ki Hong}
\affiliation{Department of
Physics, Pusan National University,
             Busan 609-735, Korea}

\date{\today}

\begin{abstract}
QCD predicts matter at high density should exhibit color superconductivity. We review briefly several pertinent properties of color superconductivity and then discuss how baryons are realized in color superconductors. Especially, we explain an attempt to describe  the color-flavor locked quark matter in terms of bosonic degrees of freedom, where the gaped quarks and Fermi sea are realized as Skyrmions, called superqualitons, and  $Q$-matter, respectively.
\end{abstract}

\maketitle

\section{Introduction}

Quantum chromodynamics (QCD) is now widely accepted as an undisputed theory of strong interactions.  The QCD prediction on how its coupling changes at different energies is thoroughly tested, and well confirmed, for the wide range of energy  from order of $1~{\rm GeV}$ to a few hundred GeV  by numerous and independent experiments.
QCD is however extremely difficult to solve, since it is highly non-linear and  strongly coupled at the same time, offering no apparant expansion parameters. So far it has precluded any analytic solutions.

One of the resason why QCD is hard to solve  is that quarks and gluons, the basic degrees of freedom of QCD,  become less relevant at low energy, where they are strongly coupled. Since the right degrees of freedom of  strong interactions at low energy are hadrons rather than quarks and gluons,  one may try to solve QCD in terms of hadrons. The Skyrme picture based on chiral Lagrangin is such an attempt~\cite{skyrme}\footnote{Holographic QCD~\cite{Sakai:2004cn} is one of the recent attempts to solve QCD directly with hadrons, especially with pions and vector mesons.}. One writes down the effective Lagrangian for the pions in powers of momentum  in accord with the QCD realization of (chiral) symmetry and then determines the couplings in the effective Lagrangian by experimental data. One interesting aspect of chiral Lagrangian is that it admits a topological solition, which is stable if one allows the so-called Skyrme term only for quartic couplings. One can also show that the topological current associated with the soliton is nothing but the baryon number current which arises from  the Wess-Zumino-Witten term~\cite{Witten:1983tw}. The baryons are therefore realized as topological solitions, known as Skyrmions in the chiral Lagrangian. The phenomenology of Skyrme Lagrangian was quite successful~\cite{Adkins:1983ya}.

Recently QCD at high density~\cite{Alford:2007xm,Hong:2007qn} has been studied intensively not only because it is relevant to dense matter,
found in compact stars like neutron stars or in heavy ion collisions, but it may shed  light on the nonperturbative behavior of QCD like chiral symmetry breaking and color confinement.  Futhermore, it is an interesting question to ask how the Skyrme picture changes as one increases baryon density, which will be addressed in this article.

The study of dense matter is ultimately related to  the properties of quarks, the basic  building blocks of atomic nuclei; how QCD and electroweak interaction of quarks behave at high baryon  density. QCD predicts  because of the asymptotic freedom a phase transition at baryon density around the QCD scale, $1/\Lambda_{\rm QCD}^3$,  that dense hadronic matter become quark matter~\cite{Collins:1974ky}. The wave function of quarks inside a nucleon overlaps with those of quarks in other nucleons as nucleons pack closely at high density, liberating quarks from nucleons.

QCD also predicts that quark matter should be color superconducting at high baryon density~\cite{Barrois:1977xd} since it is energetically prefered for quarks to form  Cooper-pairs rather than to form quark-anti-quark condensates.   Though color superconductivity has not been observed yet, one expects however to find it in the core of compact stars like neutron stars or quark stars. Finding color superconductivity will be a great challange for QCD.

\section{Color Superconducting Quark Matter}

Unlike ordinary electron superconductors,  color-superconducting quark matter has a rather rich phase structure because quarks have not only three different color charges but also come in several flavors, which makes
it extremely interesting to find color superconductors. The number of quark flavors in quark matter depends on its density because the mass gap is flavor-dependent.

At intermediate density where the strange quarks are too heavy to populate, only up and down quaks participate in Cooper-pairing. Since the color anti-triplet channel provides attraction among quarks, the quark Cooper pairs are flavor singlet but transform as anti-fundamental under ${\rm SU}(3)_c$ color:
\begin{equation}
\left<q_{Li}^q(\vec p)q_{Lj}^b(-\vec p)\right>=-\left<q_{Ri}^q(\vec p)q_{Rj}^b(-\vec p)\right>=\epsilon_{ij}\epsilon^{ab3}\Delta_2,
\end{equation}
where $i,j=1,2$ and $a,b=1,2,3$ are flavor and color indices, respectively, and $\Delta_2$ is the gap openned at the Fermi surface of two-flavor quark matter. (We will call the 3 direction in color space as blue.)
In the ground state of two flavor quark matter the Cooper pairs form condensates, breaking ${\rm SU}(3)_c$ down to its subgroup ${\rm SU}(2)_c$.  Since the Cooper pairs are flavor singlet, the ground state preserves all the global symmetries of QCD except the ${\rm U}(1)$ baryon number, which is broken down to $Z_2$ by the condensation of the Cooper pairs.  In two-flavor color superconductors five among eight gluons  are coupled to the Cooper-pairs, becoming massive due to Higgs mechansim, and the Cooper-pair gap opens at the Fermi surface of green and red quarks.

The confinement scale, $\Lambda_C$, of unbroken ${\rm SU}(2)_c$ is much smaller than the QCD scale~\cite{Rischke:2000cn} and also parametrically much smaller than the gap, $\Delta_2$.
At energy lower than $\Lambda_C$ (and  also lower than the Cooper-pair gap, $\Delta_2$), the particle spectrum of two-flavor quark matter consists of a massless Nambu-Goldstone boson accociated with broken ${\rm U}(1)_B$ and gapless up and down blue quarks, which should remain gapless as the chiral symmetry is unbroken, while four massive  gluons and gapped quarks (red and green quarks), which are fundamental under ${\rm SU}(2)_c$,  are confined in bound states like baryons or glueballs~\cite{Casalbuoni:2000cn,Ouyed:2001fv}. (The massive 8th gluon is neutral under ${\rm SU}(2)_c$ and thus decoupled from the rest of particles.)  Baryons in two-flavor color superconductors are like a heavy (blue) quarkonium made of red and green quarks.

\section{Color-Flavor Locked Quark Matter}

As nucleons pack closely together, they will eventually form quark matter. The critial density or critical chemical potential for the phase transition to quark matter  is rather difficult to estimate due to the nature of strong interactions.  While the lattice calculation for the critical temperature to form quark matter has been quite successful~\cite{Cheng:2007jq}, lattice  is of not much help at finite density due to the notorious sign problem associated with the complex measure, barring the Monte Carlo method~\cite{Hands:2007by,Hong:2002nn}. One expects, however, the phase transition at finite density presumably occurs around at the quark chemical potential, $\mu\sim\Lambda_{\rm QCD}$, solely on dimensional grounds, which corresponds to about 5 to 10 times the nuclear density, $n_0\simeq0.17~{\rm fm}^{-3}$.

In the previous section we assumed strange quarks are decoupled at intermediate density.  However, matter at density close to the critical density strange quarks are not completely decoupled as the quark chemical potential is
comparable to the strange quark mass, $m_s\simeq 100~{\rm MeV}$.
Significant fraction of quark matter is therefore composed of strange quarks. Whether strange quarks participate in Cooper-paring with up and down quarks near the critical density is still an open issue, because we do not know yet whether the pairing gap is bigger than the stress to break pairing, $m_s^2/(2\mu)$, due to the mismatch of the Fermi surfaces of pairing quarks. On the other hand at density much higher than the critical density one surely expects that all of three light flavors do participate in Cooper-pairing. In fact one can show rigorously
 that Cooper-pairs take a so-called color-flavor locked (CFL) form~\cite{Alford:1998mk} at asymptotic density~\cite{Hong:2003zq},
\begin{equation}
\left<q_{Li}^a(\vec p)q_{Lj}^b(-\vec p)\right>=-\left<q_{Ri}^a(\vec p)q_{Rj}^b(-\vec p)\right>=\Delta\,\epsilon^{ab\alpha}\epsilon_{ij\alpha}\,,
\end{equation}
where the flavor indices $i,j$ now run from 1 to 3 and  we neglected the color-sextet components, since the instanton effect is negligible.

The color-flavor locked phase of quark matter turns out to be quite stable against various stress~\cite{Alford:2003fq} and also theoretically very interesting. The particle spectrum of   CFL phase maps one-to-one  onto that of  low density (hypernuclear) hadron matter, as if there is no phase boundary between them~\cite{Schafer:1998ef}.   The chiral symmetry is spontaneously broken because the rotations of both left and right-handed quarks are locked to the same color-rotations.
If one rotates both color and flavor simultaneously, the Cooper pairs remain invariant.
The condensate of CFL Cooper-pairs also breaks the ${\rm U}(1)_{\rm em}$ electromagnetism.  Since the quarks transform under ${\rm U}(1)_{\rm em}$ as
\begin{equation}
q\mapsto e^{i\varphi\,Q_{\rm em}}\,q
\end{equation}
where $Q_{\rm em}={\rm diag}\left(2/3,-1/3,-1/3\right)$, the ${\rm U}(1)_{\rm em}$ transformation on quarks can be undone by ${\rm U}(1)_{Y}$ color  hyper-charge transformation. A linear combination of photon and hyper-charge component of gluons, $A_{\mu}^Y$, remain un-Higgsed. The modified photn of unbroken ${\rm U}(1)_{\tilde Q}$ is given as
\begin{equation}
{\tilde A}_{\mu}=A_{\mu}\cos\theta+A_{\mu}^Y\sin\theta,\quad \cos\theta=\frac{g_s}{\sqrt{e^2+g_s^2}}\,,
\end{equation}
where $e$ is the electromagnetic coupling and $g_s$ is the QCD coupling.
The symmetry breaking pattern in CFL phase is therefore given as
\begin{equation}
{\rm SU} (3)_c\times {\rm U}(1)_{\rm em}\times {\rm SU} (3)_L\times {\rm SU} (3)_R\times {\rm U} (1)_B\to {\rm SU} (3)_{c+L+R}\times Z_2\times {\rm U}(1)_{\tilde Q}\,.
\end{equation}

At high baryon density antiquarks are highly suppressed, since it takes energy bigger than the chemical potential to excite them. An effective theory of modes near the Fermi surface, called High Density Effective Theory, has been derived  by integrating out the modes far away from the Fermi surface~\cite{Hong:1998tn,Schafer:2003jn}. The Dirac mass term, which breaks the chiral symmetry explicitly, is suppressed dense medium and gives  mass operators as, once antiquarks are integrated out,
\begin{equation}
m\bar qq=\frac{mm^{\dagger}}{2\mu}q_+^{\dagger}q_++\frac{mm^T}{4\mu^2}\left(q^{\dagger}_+\Delta\, q_{c+}+{\rm h.c.}\right)+\cdots\,,
\end{equation}
where $q_+$ denotes quarks near the Fermi surface and $q_{c+}$ their chage-conjugate fields~\cite{Hong:2000ng}. Therefore, mass of pseudo NG bosons becomes suppressed in dense medium as $m^2/(2\mu)$.
The particle spectrum of CFL phase consists of 8 pseudo Nambu-Goldstone (NG) bossons of mass $m^2/(2\mu)$  and one massless Nambu-Goldstone boson, corresponding to the baryon superfluid, and 8 massive gluons, and 9 massive quarks. Under the unbroken  global symmetry, ${\rm SU}(3)_{c+L+R}\times Z_2$, the particles transform as in Table~\ref{spectrum} \footnote{The baryon number is spontaneously broken by a condensate of Cooper pairs, which carry $B=2/3$. So, it is defined modulo $2/3$.}
\begin{center}
\begin{table}[hbp]
\begin{tabular}{|c|c|c|c|c|c|}
\hline
Particles & Spin & Mass &${\rm SU}(3)_{c+L+R}$ & ${\rm U}(1)_B$ & $Z_2$\\
\hline
NG bosons & 0 &${\rm O}(\frac{m^2}{2\mu})$ & $8\oplus1$ &0&$+1$\\
\hline
Gapped quarks &1/2 &$\Delta$ &8 & $\frac13$ &$-1$\\
\hline
Gapped quarks &1/2 &$2\Delta$ &1 & $\frac13$ &$-1$\\
\hline
Gluons & 1 & ${\rm O}(g_s\Delta)$ &8&0&+1\\
\hline
\end{tabular}
\caption{particle spectrum of CFL phase}
\label{spectrum}
\end{table}
\end{center}

The ground state of the CFL phase is nothing but the Fermi sea with gap opened at  Fermi surfaces of all nine quarks by
Cooper-pairing; the octet under ${\rm SU}(3)_{c+L+R}$ has a gap $\Delta$ while the singlet has $2\Delta$.
The collective excitations of Cooper-pairs are possible without any energy gap at arbitrarily low energy, exhibiting superflows of mass and color charges~\footnote{The Coper-pair breaks ${\rm U}(1)$ electromagnetism but leaves ${\rm U}(1)_{\tilde Q}$ unbroken, a linear combination of  ${\rm U}(1)$ electromagnetism and ${\rm U}(1)_Y$ color-hypercharge .  Therefore CFL quark matter will have a finite resistance for ${\tilde Q}$ charges.}
To describe the low-energy excitations of the CFL quark matter, we introduce composite
(diquark) fields $\phi_L$ and $\phi_R$ as
\begin{equation}
{\phi_{L(R)}}_{ai}(x)\equiv\lim_{y\to x}{\left|x-y\right|^{\gamma_m}
\over\kappa}\,\epsilon_{abc}\epsilon_{ijk}
q^{bj}_{L(R)}(-\vec v_F,x)q^{ck}_{L(R)}(\vec v_F,y),
\label{diquark}
\end{equation}
where $\gamma_m$  is the anomalous dimension of the
diquark fields and $q(\vec v_F,x)$ denotes a quark field with
momentum close to a Fermi momentum $p_F$~\cite{Hong:1998tn}.  A dimensional quantity $\kappa$ is introduced in (\ref{diquark}) so that the expectation value of diquark fields in the ground state becomes identity,
\begin{equation}
\left<\phi_L\right>=-\left<\phi_R\right>=I\,.
\end{equation}
Under the color and flavor symmetry the diquark fields transform as
\begin{equation}
\phi_L\mapsto g_c^T\,\phi_L\,g_L\,,\quad \phi_R\mapsto g_c^T\,\phi_R\,g_R\,,\quad g \in{\rm SU}(3)\,.
\end{equation}
The low-lying excitations of condensates are then described by following unitary matrices,
\begin{equation}
U_L(x)=g_c^T(x)g_L(x),\quad U_R(x)=g_c^T(x)g_R(x)\,,
\end{equation}
which may be parametrized by Nambu-Goldstone fields as
\begin{equation}
U_{L(R)}(x)=\exp\left(2i\Pi_{L(R)}^A(x)T^A/F_{\pi}\right)\,,
\end{equation}
where $T^A$ are the ${\rm SU}(3)$ generators, normalized as ${\rm Tr}\,T^AT^B=\frac12\delta^{AB}$. The parity-even combination of the Nambu-Goldstone bosons, constituting the longitudinal components of gluons, generates the color supercurrents in quark matter, while the parity-odd combination becomes pseudo Nambu-Goldstone bosons, neutral under color, which can be written as
\begin{equation}
\Sigma_i^j(x)\equiv U_{Lai}(x){U^*}_R^{aj}(x)=\exp\left(2i\,\Pi^A(x)T^A/F\right)\,,
\end{equation}
where $\Pi^A(x)=\Pi^A_L(x)-\Pi^A_R(x)$ are correlated excitatons of $\phi_L$ and $\phi_R$, having same quantum numbers as pions, kaons, and eta in hadronic phase.

Expanding in powers of derivatives, the low-energy effective Lagrangian density for the
(colored) Nambu-Goldstone bosons  is given as
\begin{eqnarray}
{\cal L}_0=\frac{F^2}{4}\,{\rm Tr}\,\left(\partial_0U_L\partial_0U_L^{\dagger}-v^2\,\vec\nabla U_L\cdot\vec\nabla U_L^{\dagger}\right)+n_L{\cal L}_{\rm WZW}+\left(L\leftrightarrow R\right)+\cdots\,,
\label{plain}
\end{eqnarray}
where $v=1/\sqrt{3}$ is the speed of Nambu-Goldstone bosons in medium and the Wess-Zumino-Witten (WZW) term ${\cal L}_{\rm WZW}$ is added to correctly reproduce the  symmetries of dense QCD. The colored NG bosons will couple to (massive) gluons through minimal coupling, replacing the ordinary derivatives with covariant derivatives, $D=\partial +ig_sA$, which amounts to adding to the effective Lagrangian, (\ref{plain}), the minimal  gauge coupling and the gluon mass
\begin{equation}
{\cal L}_1=-g_sA_{\mu}J^{\mu}-m_g^2\,{\rm Tr}\,A_{\mu}^2
\end{equation}
and also replacing with covariant ones  the plain derivatives in Wess-Zumino-Witten term to reproduce the anomalous coupling of NG bosons with vector mesons.

The coefficient of the WZW term in the effective Lagrangian should be chosen to match the global anomalies of microscopic theory. For instance the ${\rm SU}(3)_L$  anomaly is given at the quark level as
\begin{equation}
\partial J_{L\mu}^A=\frac{{\tilde e}^2}{32\pi^2}\,{\rm Tr}\left(T^A{\tilde Q}^2\right)\,\epsilon_{\mu\nu\rho\sigma}{\tilde F}^{\mu\nu}{\tilde F}^{\rho\sigma}\,,
\label{anomaly}
\end{equation}
where ${\tilde F}$ is the field strength tensor of the modified photon. On the other hand the  WZW term contains a term, if one gauges ${\rm U}(1)_{\tilde Q}$,
\begin{equation}
{\cal L}_{WZW}\ni \frac{n_L{\tilde e}^2}{64\pi^2F}\Pi^0 \epsilon_{\mu\nu\rho\sigma}{\tilde F}^{\mu\nu}{\tilde F}^{\rho\sigma}\,,
\end{equation}
which agrees with (\ref{anomaly}) if $n_L=1$ since ${\rm Tr}\left(T^3{\tilde Q}^2\right)=1/2$~\cite{Hong:1999dk}.
Similarly, one can show that $n_R=1$.

\section{Superqualitons and gaped quarks}
In vacuum chiral symmetry breaking occurs due to the condensation of quark-antiquark bilinear at strong coupling. The coefficients of operators in the chiral effective Lagrangian therefore contain the physics of strong dynamics and are hence very difficult to calculate directly from QCD. However, the chiral symmetry breaking in the color-flavor locked phase of quark matter occurs even at asympotic density where the QCD coupling is extremely small, because it is due to the Cooper-paring of quarks near the Fermi surface which can occur at arbitrarily weak attraction due to Cooper theorem.

The coefficients of operators in the low-energy effective Lagrangian of CFL matter are calculable at asymptotic density, using perturbation, called hard dense-loop approximation, which appropriately incorporates the medium effects.  Similarly, the CFL gap, which characterizes the properties of CFL matter, can be also calculated precisely, using perturbation. However at not-so-high density where  CFL matter is strongly coupled we do not have  well-developed tools to study neither the gap nor  the low-energy constants of the effective theory, as in the vacuum QCD. To study the proerties of CFL matter at intermediate density the Skyrme's idea may be useful, which correctly captures the large ${\rm N}_c$ behavior of baryons as topological solitons made of pions. In the case of CFL matter where quarks are deconfined, though gapped, the topological solitons made out of (colored) NG bosons, called superqualitons, should be identified as gapped quarks, similar to Kaplan's qualiton~\cite{Kaplan:1989gs} which realizes the constitituent quarks inside nucleons. In this section we study the CFL gap of strongly interacting quark matter \'a~la Skyrme~\cite{Hong:1999dk}.

We first note that gaped quarks of each chirality should be treated independently, since  Cooper-pairing occurs between quarks with same chirality. (We will concentrate on the left-handed gaped quark. But, the argument below applies equally to the right-handed gaped quark.) The manifold of NG bosons, $\Pi_L$, associated  the condensation of Cooper pairs of left-handed quarks
has a notrivial third homotopy,
\begin{equation}
\Pi_3\left({\rm SU}(3)_c\times {\rm SU}(3)_L/{\rm SU}(3)_{c+L}\right)=Z_n\,,
\end{equation}
and thus the low-energy effective Lagrangian of $\Pi_L$ admits a topological soliton associated with a topological current,
\begin{equation}
J_L^{\mu}=\frac{1}{24\pi^2}\epsilon^{\mu\nu\rho\sigma}\,{\rm Tr}\,U_L^{-1}\partial_{\nu}U_L U_L^{-1}\partial_{\rho}U_L U_L^{-1}\partial_{\sigma}U_L\,,
\end{equation}
whose charge counts the number of winding of the map $U_L$ from $S^3$, the boundary of space at infinity to the manifold of NG bosons.
Since the sigma model description of the ${\rm SU}(3)_L$ quark current $J_{L\mu}^A=\bar q_{L+} T^A\gamma_{\mu}q_{L+}$ contains an anomalous piece from the WZW term,
\begin{equation}
J_{L\mu}^A\ni\,\frac{1}{24\pi^2}\epsilon^{\mu\nu\rho\sigma}\,{\rm Tr}\,T^AU_L^{-1}\partial_{\nu}U_L U_L^{-1}\partial_{\rho}U_L U_L^{-1}\partial_{\sigma}U_L\,
\end{equation}
the topological current should be interepreted as the (left-handed) quark number current, $J_{L\mu}= \bar q_{L+}\gamma_{\mu}q_{L+}$ and the soliton of unit winding number should be identified as (left-handed) gaped quark, carrying a baryon number $1/3$ as $n_L=1$ rather 3 in the case of vacuum QCD.

Once we identify the soliton as a gaped quark, it is staightforward to estimate the magnitude of the gap as a function of the low energy constants like the NG boson decay constant, $F$, or the QCD coupling, as the soliton is stabilzed by the Coulomb repulsion due to color charges at the core.  Following Skyrme~\cite{skyrme} we seek a static
configuration for the field $U_L$ in $SU(3)$ by embedding an
$SU(2)$ hedgehog in color-flavor in $SU(3)$, with
\begin{equation}
{U_L}_c(x)=e^{i\vec\tau\cdot \hat r\theta(r)},\quad U_R=0,
\quad A^Y_0=\omega(r),\quad {\rm all~~ other}~~A_{\mu}^A=0,
\label{profile}
\end{equation}
where $\tau$'s are Pauli matrices. The radial function $\theta(r)$
is monotonous and satisfies
\begin{equation}
\theta(0)=\pi,\quad\theta(\infty)=0
\label{bound}
\end{equation}
for a soliton of winding number one.
(Note that we can also look for a right-handed soliton by switching
off the $U_L$ field. The solution should be identical because
QCD is invariant under parity.) This configuration
has only non-vanishing color charge in the color-hypercharge $Y$ direction
\begin{equation}
J_{0}^Y={\sin^2\theta~\theta^{\prime}\over 2\pi r^2}
\end{equation}
and all others are zero.
As shown in~\cite{Hong:1989aa}, the energy of the configuration
(\ref{profile}) is given as
\begin{equation}
E[\omega,\theta]=\int 4\pi r^2{\rm d}r\left[
-{1\over2}{\omega^{\prime}}^2+F^2\left({\theta^{\prime}}^2
+2{\sin\theta^2\over r^2}\right)+{g_s\over 2\pi^2}
{\omega\over r^2}\sin^2\theta~\theta^{\prime}
\right]\,
\end{equation}
and the total charge within a radius $r$ is
\begin{equation}
Q^Y(r)=g_s\int_0^r {\rm Tr}~YJ^Y_0(r^{\prime})4\pi
{r^{\prime}}^2{\rm d}r^{\prime}
=-g_s\left({\theta(r)-\sin\theta(r)\cos\theta(r)-\pi\over\pi}
\right).
\end{equation}
Using the Gauss law with screened charge density,
we can trade  $\omega$ in terms of $\theta(r)$,
\begin{equation}
\omega^{\prime}={Q^Y(r)\over 4\pi r^2}e^{-m_Er},
\end{equation}
where $m_E=\sqrt{6\alpha_s/\pi}\,\mu$ is the electric screening mass
for the gluons.
Hence, the energy functional simplifies to
\begin{equation}
E[\theta]=\int_0^{\infty}{\cal E}(r) \, {\rm d}r =
\int_0^{\infty}4\pi{\rm d}r\left[
F^2\left(r^2{\theta^{\prime}}^2+2\sin^2\theta\right)
+{\alpha_s\over2\pi}\left({\theta-\sin\theta\cos\theta-\pi
\over2r}\right)^2e^{-2m_Er}
\right],
\label{mass}
\end{equation}
where $\alpha_s=g_s^2/(4\pi)$.
The squared size of the superqualiton is $R_S^2= \left<r^2\right>$
where the averaging is made using the (weight) density
${\cal E}(r)$.

The equation of motion for the superqualiton profile $\theta (r)$ is
\be
\left(r^2\theta'\right)' ={\rm sin}2\theta +
\frac {\alpha_s}{4\pi}\frac{e^{-2m_Er}}{(F\, r)^2}\sin^2\theta
\left(\pi -\theta+\frac 12 {\rm sin}2\theta \right)
\label{mot}
\ee
subject to the boundary conditions (\ref{bound}).
We solve the equation Eq.~(\ref{mot})
numerically for several values of $m_E$ and $\alpha_s$.
The profile function of the superqualiton for $m_E=20F$
and $\alpha_s=1$ is shown in Fig.~1.
\begin{figure}
\vskip 0.2in
\centerline{\epsfbox{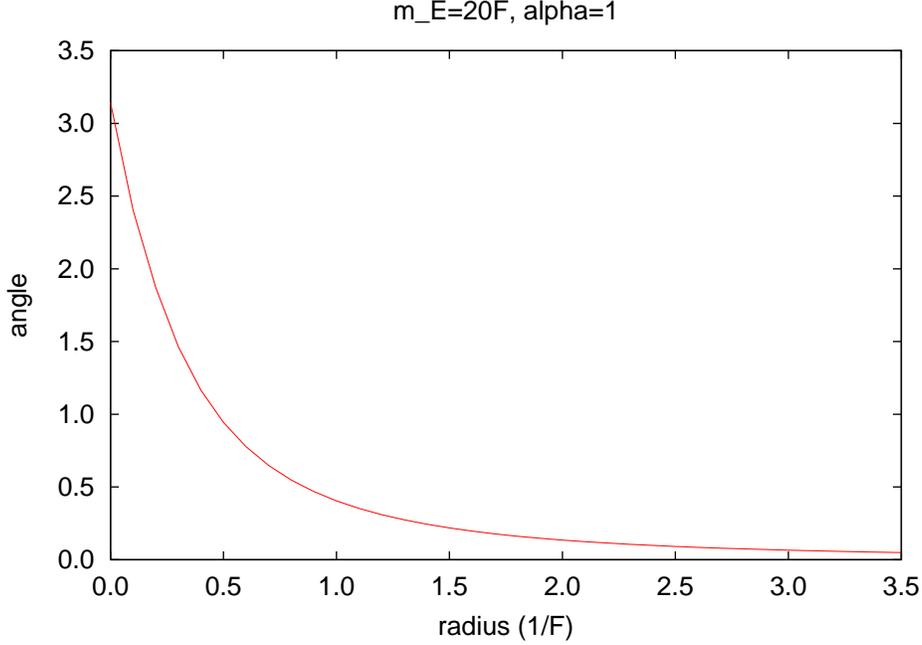}}
\vskip 0.2in
\caption{The qualiton profile for $m_E=20F$ and $\alpha_s=1$.}
\end{figure}
For $m_E/F=1,10,100$ and $\alpha_s=1$ we find the soliton mass
$M_S=2.41, ~2.08,~2.07\times 4\pi F$ and
$R_S=1.36,~1.35,~1.347~F^{-1}$,
respectively. By varying the coupling for a
fixed screening mass, $m_E=F$,
we find $R_s=1.25,~1.30,~1.58/F$ for
$\alpha_s=0.1,~1,~10$, respectively, showing that
the soliton gets bigger for the stronger coupling, since
the color-electric force, which balances the kinetic energy
of the soliton, is more repulsive~\cite{Hong:1989aa}.

To access the quantum numbers and the spectrum of the superqualiton,
we note as usual that for any static solution to the equations
of motion,
one can generate another solution by a rigid $SU(3)$ rotation,
\begin{equation}
U(x)\mapsto A U(x)A^{-1},\quad A\in SU(3).
\end{equation}
The matrix $A$ corresponds to the zero modes of the superqualiton.
Note that two $SU(3)$ matrices are equivalent if they differ by
a matrix $h\in U(1)\subset SU(3)$ that commutes with $SU(2)$
generated by $\vec \tau\otimes I$.
The Lagrangian for the zero modes is given by substituting
$U(\vec x,t)=A(t)U_c(\vec x)A(t)^{-1}$~\cite{bal85}. Hence,
\begin{equation}
L[A]=-M_S+{\frac12}I_{\alpha\beta}{\rm Tr}~T^{\alpha}A^{-1}
\dot{A}{\rm Tr}~T^{\beta}A^{-1}{\dot A}
-i{1\over 2}{\rm Tr}~YA^{-1}{\dot A},
\end{equation}
where $I_{\alpha\beta}$ is an invariant tensor on
${\cal M}=SU(3)/U(1)$ and the hypercharge $Y$ is
\begin{equation}
Y=\frac13
\begin{pmatrix}
1&0&0\\
0&1&0\\
0 & 0 &-2
\end{pmatrix}\,.
\end{equation}
Under the transformation $A(t)\mapsto A(t)h(t)$ with
$h\in U(1)_Y$
\begin{equation}
L\mapsto L-\frac{i}{2}{\rm Tr}~Yh^{-1}\dot h.
\end{equation}
Therefore, if we rotate adiabatically the soliton by $\theta$ in the
hypercharge space in $SU(3)$, $h=\exp(iY\theta)$,
for time $T\to\infty$, then the wave
function of the soliton changes by a phase in the semiclassical limit;
\begin{equation}
\psi(T)\sim e^{i\int {\rm d}tL}\psi(0)=e^{i\theta/3}\psi(0),
\label{wfn}
\end{equation}
where we neglected the irrelevant phase $-M_ST$ due to the rest mass
energy. The simplest and lowest energy configuration that satisfies
Eq.~(\ref{wfn}) is the fundamental representation of $SU(3)$.
Similary, under a spatial (adiabatic) rotation by $\theta$
around the $z$ axis, $h=\exp(i\tau^3\theta)$,
the phase of the wave function changes by $\theta/2$.
Therefore, the ground state of the soliton is
a spin-half particle transforming under the fundamental representation
of both the flavor and the color group, which leads us to conclude that
the soliton is a gaped left-handed (or right-handed)
quark in the CFL phase.

\section{Bosonization of QCD at high density}
The gap in superconductors can be estimated by measuring the energy needed to excite a pair of particle and hole, breaking a Cooper-pair. If one decreases the total number of particles by $\delta N$,  creating holes in Fermi Sea, the thermodynamic potential (or total energy at zero temperature) of ground state is reduced  by $\delta {E}=\mu\delta N$.  Therefore the gap in the superqualtiton should be defined as~\footnote{One needs energy, twice of the gap, to break a Cooper pair.}
\begin{equation}
\Delta=\frac12\left(M_S-\delta E\right)\,,
\end{equation}
where $M_S$ is mass of soliton, calculated from~(\ref{mass}).

Quark matter with finite baryon number is described
by QCD with a chemical potential, which restricts the system
to have a fixed baryon number on average;
\begin{equation}
{\cal L}={\cal L}_{\rm QCD}-\mu \sum_{i=u,d,s}\bar q_i\gamma^0q_i,
\end{equation}
where $\bar q_i\gamma^0q_i$ is
the quark number density of the $i$-th flavor.
The ground state in the CFL phase is nothing but
the Fermi sea where all quarks are Cooper-paired.
Equivalently,
this system can be described in term of bosonic
degrees of freedom, namely pions and kaons, which are small
fluctuations of Cooper pairs~\cite{Hong:2000ff}.

As the baryon number (or the quark number) is conserved,
though spontaneously broken,
the ground state in the bosonic description should
have the same baryon (or quark) number as the ground state
in the fermionic description.
Under the $U(1)_Q$ quark number symmetry,
the bosonic fields transform as
\begin{equation}
U_{L,R}\mapsto e^{i\theta Q}U_{L,R}e^{-i\theta Q}=e^{2i\theta}U_{L,R},
\end{equation}
where $Q$ is the quark number operator,
given in the bosonic description as
\begin{equation}
Q=i\int {\rm d}^3x~{F^2\over4}
{\rm Tr}\left[U_L^{\dagger}\partial_tU_L-\partial_tU_L^{\dagger}U_L
+\left(L\leftrightarrow R\right)\right].
\end{equation}

The energy in the bosonic description is
\begin{equation}
E=\int{\rm d}^3x{F^2\over 4}
{\rm Tr}\left[\left|\partial_tU_L\right|^2
+\left|\vec\nabla U_L\right|^2
+\left(L\leftrightarrow R\right)\right]+E_m+\delta E,
\end{equation}
where $E_m$ is the energy due to meson mass and $\delta E$ is the
energy coming from the higher derivative terms. Assuming
the meson mass energy is positive and $E_m +\delta E\ge0$, which
is reasonable because $\Delta/F\ll1$,
we can take, dropping the positive terms due to the spatial derivative,
\begin{equation}
E\ge \int{\rm d}^3x{F^2\over 4}
{\rm Tr}\left[\left|\partial_tU_L\right|^2
+\left(L\leftrightarrow R\right)\right](\equiv E_Q).
\end{equation}
Since for any number $\alpha$
\begin{equation}
\int{\rm d}^3 x~{\rm Tr}\left[\left|U_L+\alpha i\partial_tU_L
\right|^2+\left(L\leftrightarrow R\right)\right]\ge0,
\end{equation}
we get a following Schwartz inequality,
\begin{equation}
Q^2\le  I\,E_Q,
\label{bound}
\end{equation}
where we defined
\begin{equation}
I={F^2\over 4}\int{\rm d}^3x\,{\rm Tr}\left[U_LU_L^{\dagger}
+\left(L\leftrightarrow R\right)\right].
\end{equation}
Note that the lower bound in Eq.~(\ref{bound})
is saturated for $E_Q=\omega Q$ or
\begin{equation}
U_{L,R}=e^{i\omega t} \quad{\rm with}\quad \omega={Q\over I}.
\end{equation}
The ground state of the color superconductor, which has the lowest
energy for a given quark number $Q$, is nothing but a so-called
$Q$-matter, or the interior of a very large
$Q$-ball~\cite{Coleman:1985ki,Hong:1998ur}. Since in the fermionic
description the system has the quark number $Q=\mu^3/\pi^2\int{\rm
d}^3x=\mu^3/\pi^2\cdot I/F^2$, we find, using
$F\simeq0.209\mu$~\cite{Son:2000cm},
\begin{equation}
\omega={1\over\pi^2}\left({\mu\over F}\right)^3 F
\simeq2.32\mu,
\label{fpi}
\end{equation}
which is numerically very close to
$4\pi F$. 
The ground state of the system in the bosonic description is a
$Q$-matter whose energy per unit quark number is $\omega$.

Since, reducing the quark number of
the $Q$-matter by one, the minimum energy we gain by creating a hole in Fermi sea is $\delta E=\omega$ and therefore the energy cost to
create a gaped quark from the ground state in the bosonic
description is
\begin{equation}
2\Delta=M_S-\omega,
\label{deltae}
\end{equation}
where $M_S$ is the energy of the superqualiton configuration in~(\ref{mass}). From~(\ref{deltae}) we can estimate the CFL gap of strongly interacting quark matter~\cite{Hong:2000ff}.

\section{Conclusion}
Solving QCD is an outstanding problem in physics. We review an attempt to solve three-flavor QCD at finite density in term of pions, kaons, and eta that occur as collective excitations of condensed Cooper-pairs, following Skyrme's idea that was applied to strong interactions.  This attempt is promising and conceptually beautiful, since it deals with the correct degrees of freedom at low energy.
The ground state of color-flavor locked phase is realized as a $Q$-matter, a collective excitation of Nambu-Golstone bosons, carrying a fixed baryon number. The gaped quarks are realized as topological solitions, made of NG bosons, similar to Kaplan's qualiton picture of constituent quarks.

The Skyrme's picture on baryons is used to estimate the color-flavor locked gap in strongly interacting quark matter, where perturbation fails,  after correctly identifying the ground state of color-superconducting quark matter.

\vskip 0.2in

\acknowledgments
The author thanks M. Rho for the invitation to contribute to this volume. He is also grateful to  S.~T. Hong, Y.~J. Park, M.~Rho, and I.~Zahed for the collaborations upon which this review is based on.
This work was supported in part  by KOSEF
Basic Research Program with the grant No. R01-2006-000-10912-0 and
also by the Korea Research Foundation Grant funded by
the Korean Government (MOEHRD, Basic Research Promotion Fund) (KRF-2007-314-
C00052)

\end{document}